\documentstyle[12pt]{article}

\newcommand{\beq}[1]{\begin{equation}\label{#1}}
\newcommand{\eeq}{\end{equation}}
\newcommand{\bear}[1]{\begin{eqnarray}\label{#1}}
\newcommand{\ear}{\end{eqnarray}}
\newcommand{\nn}{\nonumber}

\catcode`\@=11 \@addtoreset{equation}{section}\catcode`\@=12
\newcommand{\be}{\begin{equation}}
\newcommand{\ee}{\end{equation}}
\newcommand{\ba}{\begin{eqnarray}}
\newcommand{\ea}{\end{eqnarray}}

\newcommand{\np}{ {\newpage } }
\newcommand{\0}{ {\bf 0 } }
\newcommand{\1}{ {\bf 1 } }

\newcommand{\R}{ {\bf R} }
\newcommand{\C}{ {\bf C} }
\newcommand{\G}{ {\bf G} }

\newcommand{\eps}{ \varepsilon }

\newcommand{\p}{\partial}
\newcommand{\btd}{\bigtriangledown}

\newcommand{\const}{\mathop{\rm const}\nolimits}

\begin{document}

\begin{center}    \bf \large

On supersymmetric solutions in $D = 11$ supergravity \\ with
Ricci-flat internal spaces

\end{center}

\vspace{1.03truecm}

\bigskip

\centerline{\bf V. D. Ivashchuk }

\vspace{0.96truecm}

\centerline{Centre for Gravitation and Fundamental Metrology}
\centerline{VNIIMS, 3-1 M. Ulyanovoy St.}
\centerline{Moscow, 117313, Russia, and}
\centerline{Institute of Gravitation and Cosmology,}
\centerline{Peoples' Friendship University of Russia,}
\centerline{6 Miklukho-Maklaya Str., Moscow 117198, Russia}

\centerline{e-mail: ivas@rgs.phys.msu.su}

\begin{abstract}

An introduction to supersymmetric  (SUSY) solutions
defined on  the product of Ricci-flat spaces
in $D= 11$ supergravity is presented.
The paper contains some background information: 
(i) decomposition relations for SUSY equations and 
(ii) $2^{-k}$-splitting theorem
that explains the appearance of $N = 2^{-k}$ fractional supersymmetries.
Examples of $M2$ and $M5$ branes on the product
of two Ricci-flat spaces are considered and formulae
for (fractional) numbers of unbroken SUSY are obtained.

\end{abstract}

\np

\section{\bf Introduction}
\setcounter{equation}{0}

In this paper we start "SUSY investigations"
of solutions with (intersecting composite non-localized)
$p$-branes from \cite{IM0}-\cite{IM3}).
These solutions have block-diagonal metrics defined on the product of
Ricci-flat spaces $M_{0}  \times M_{1} \times \ldots \times M_{n}$
and are governed by harmonic functions on $M_0$.
Thus, we are interesting in a subclass of SUSY solutions
>from \cite{IM0}-\cite{IM3}), when certain supergravity theories
are considered. Here we consider the case of $D= 11$ supergravity
\cite{CJS}.

We note that for flat internal spaces and $M_0$ the
SUSY solutions were considered intensively in numerous publications
(see, e.g. \cite{DKL}-\cite{Ts} and references therein).

Recently certain  SUSY solutions in $D= 10, 11$ supergravities
with several internal Ricci-flat internal spaces were considered
\cite{DLPS}-\cite{MPl}. Some of them may be obtained by a simple
replacing of flat metrics by Ricci-flat ones.
We note that major part of these exact solutions
(regardless to SUSY properties) are not new ones but are special
cases of those obtained before (see \cite{IM0}-\cite{IM3},
and references therein). (For example, the 
magnetic $5$-brane solution from \cite{K1} 
with $N =1/4$ SUSY is a special case of solutions
>from \cite{IM0,IM01,IM} etc).

Here we suggest a more general approach for investigation
of the solutions to  "Killing-like"  SUSY equations
in the backgrounds with block-diagonal metric
defined on  the product of Ricci-flat spaces
$M_{0}  \times M_{1} \times \ldots \times M_{n}$
with arbitrary (though restricted) $n$.
The $4$-form  is (tacitly) assumed to be a sum of
$p$-brane  monoms of magnetic and electric types.

The  paper contains some background treatment:
({\bf i})  decomposition relation for
spin connection and  matrix-valued
covector fields that appear in the SUSY equation;
({\bf ii}) $2^{-k}$-splitting theorem for
$k$ commuting linear operators.

Here we consider the simplest examples of $M2$ and $M5$ branes
defined on the product  of two Ricci-flat spaces and obtain
formulae  for fractional number of SUSY.
We also consider the simplest $M2 \cap M5$-configuration
(defined on the product of flat spaces) to
show how the $2^{-k}$-splitting theorem works.

\section{\bf Basic notations}

Now we describe the basic notations (in arbitrary dimension $D$).

\subsection{Product of manifolds}

Here we consider the manifolds
\beq{0.10}
M = M_{0}  \times M_{1} \times \ldots \times M_{n},
\eeq
with the metrics
\beq{0.11}
g= e^{2{\gamma}(x)} \hat{g}^0  +
\sum_{i=1}^{n} e^{2\phi^i(x)} \hat{g}^i ,
\eeq
where $g^0  = g^0 _{\mu \nu}(x) dx^{\mu} \otimes dx^{\nu}$
is a metric on the manifold $M_{0}$
and $g^i  = g^{i}_{m_{i} n_{i}}(y_i) dy_i^{m_{i}} \otimes dy_i^{n_{i}}$
is a metric on $M_{i}$, $i = 1,\ldots, n$.

Here $\hat{g}^{i} = p_{i}^{*} g^{i}$ is the
pullback of the metric $g^{i}$  to the manifold  $M$ by the
canonical projection: $p_{i} : M \rightarrow  M_{i}$,
$i = 0,\ldots, n$.

The functions $\gamma, \phi^{i} : M_0 \rightarrow \R$ are smooth.
We denote $d_{\nu} = {\rm dim} M_{\nu}$; $\nu = 0, \ldots, n$;
$D = \sum_{\nu = 0}^{n} d_{\nu}$.
We put any manifold $M_{\nu}$, $\nu = 0,\ldots, n$, to be oriented
and connected.
Then the volume $d_i$-form
\beq{0.14}
\tau_i  \equiv \sqrt{|g^i(y_i)|}
\ dy_i^{1} \wedge \ldots \wedge dy_i^{d_i},
\eeq
and signature parameter
\beq{0.15}
\varepsilon(i)  \equiv {\rm sign}( \det (g^i_{m_i n_i})) = \pm 1
\eeq
are correctly defined for all $i=1,\ldots,n$.
Let $\Omega = \Omega(n)$  be a set of all non-empty
subsets of $\{ 1, \ldots,n \}$ ($|\Omega| = 2^n - 1$).
For any $I = \{ i_1, \ldots, i_k \} \in \Omega$, $i_1 < \ldots < i_k$,
we denote
\bear{0.16}
\tau(I) \equiv \hat{\tau}_{i_1}  \wedge \ldots \wedge \hat{\tau}_{i_k},  \\
 \label{0.17}
\eps(I) \equiv \eps(i_1) \ldots \eps(i_k),  \\
\label{0.19}
d(I) \equiv  \sum_{i \in I} d_i.
\ear
Here $\hat{\tau}_{i} = p_{i}^{*} \hat{\tau}_{i}$ is the
pullback of the form $\tau_i$  to the manifold  $M$ by the
canonical projection: $p_{i} : M \rightarrow  M_{i}$, $i = 1,\ldots, n$.

\subsection{Diagonalization of metric}

For the metric $g  = g_{M N}(x) dx^{M} \otimes dx^{N}$ from (\ref{0.11}),
$M,N = 0, \ldots, D-1$, defined on the manifold (\ref{0.10}),
we define the diagonalizing $D$-bein $e^A = e^A_{\ \ M} dx^{M}$
\beq{0.20}
g_{MN} = \eta_{A B} e^A_{\ \ M} e^B_{\ \ N},
\qquad \eta_{AB} = \eta^{AB} = \eta_A \delta_{AB},
\eeq
$\eta_A = \pm 1$; $A,B = 0, \ldots, D-1$.

We choose the following frame vectors
\beq{0.21}
(e^A_{\ \ M}) = {\rm diag}(e^{\gamma} e^{(0) a}_{\ \ \ \ \mu},
e^{ \phi^1} e^{(1) a_1}_{\ \ \ \ m_1}, \ldots,
e^{\phi^n} e^{(n) a_n}_{\ \ \ \ m_n}),
\eeq
where
\beq{0.22}
g^{0}_{\mu \nu} = \eta^{(0)}_{ab}
e^{(0) a}_{\ \ \ \ \mu} e^{(0) b}_{\ \ \ \ \nu},
\qquad
g^{i}_{m_i n_i} = \eta^{(i)}_{a_i b_i}
e^{(i) a_i}_{\ \ \ \ m_i} e^{(i) b_i}_{\ \ \ \ n_i},
\eeq
$i = 1, \ldots, n$, and
\beq{0.23}
(\eta_{AB}) = {\rm diag}(\eta^{(0)}_{ab},
\eta^{(1)}_{a_1 b_1}, \ldots, \eta^{(n)}_{a_n b_n}).
\eeq
For $(e^M_{\ \ A})  = (e^A_{\ \ M})^{-1}$ we get
\beq{0.24}
(e^M_{\ \ A}) = {\rm diag}(e^{- \gamma} e^{(0) \mu}_{\ \ \ \ a},
e^{- \phi^1} e^{(1) m_1}_{\ \ \ \ a_1}, \ldots,
e^{- \phi^n} e^{(n) m_n}_{\ \ \ \ a_n}),
\eeq
where $(e^{(0) \mu}_{\ \ \ \ a}) = (e^{(0) a}_{\ \ \ \ \mu})^{-1}$,
$(e^{(i) m_i}_{\ \ \ \ a_i}) = (e^{(i) a_i}_{\ \ \ \ n_i})^{-1}$,
$i = 1, \ldots, n$.

{\bf Indices}.
For indices we also use an alternative  numbering:
$A = (a,a_1, \ldots, a_n)$, $B = (b,b_1, \ldots, b_n)$,
where $a,b = 1_0, \ldots, (d_0)_0$; $a_1,b_1 = 1_1, \ldots, (d_1)_1$;
...; $a_n,b_n = 1_n, \ldots, (d_n)_n$; and
$M = (\mu,m_1, \ldots, m_n)$, $N = (\nu,n_1, \ldots, n_n)$, where $\mu,\nu = 1_0, \ldots, (d_0)_0$; $m_1,n_1 = 1_1, \ldots, (d_1)_1$;
...; $m_n,n_n = 1_n, \ldots, (d_n)_n$.

\subsection{Gamma-matrices}

In what follows $\hat{\Gamma}_A$ are "frame"
$\Gamma$-matrices  satisfying
\beq{0.25}
\hat{\Gamma}_A \hat{\Gamma}_B + \hat{\Gamma}_B \hat{\Gamma}_A
= 2 \eta_{AB} \1,
\eeq
$A,B = 0, \ldots, D-1$. Here $\1 = \1_D$ is unit $D \times D$ matrix.
We also use "world"  $\Gamma$-matrices
\beq{0.26}
\Gamma_M = e^A_{\ \ M} \hat{\Gamma}_A, \qquad
\Gamma_M \Gamma_N + \Gamma_N \Gamma_M = 2 g_{MN} \1,
\eeq
$M,N = 0, \ldots, D-1$, and the matrices with upper indices:
$\hat{\Gamma}^A=  \eta^{AB} \hat{\Gamma}_B$  and
$\Gamma^M = g^{MN} \Gamma_N$.

\subsection{Spin connection}

Here we use the standard definition for the spin connection
\beq{0.27}
\omega^{A}_{\ \ BM} = \omega^{A}_{\ \ BM}(e,\eta)
= e^A_{\ \ N}  \btd_M[g(e,\eta)] e^N_{\ \ B},
\eeq
where the covariant derivative $\btd_M[g]$ corresponds
to the metric $g = g(e,\eta)$ from (\ref{0.20}).
The spinorial covariant derivative
reads
\beq{0.28}
D_M = \p_M +  \frac{1}{4} \omega_{A B M} \hat{\Gamma}^A \hat{\Gamma}^B,
\eeq
where $\omega_{A B M} = \eta_{AA'}\omega^{A'}_{\ \ BM}$.

The non-zero components of the spin connection
(\ref{0.27}) in the frame (\ref{0.21}) read
\bear{0.29}
\omega^{a}_{\ \ b \mu} =
\omega^{a}_{\ \ b \mu}(e^{(0)},\eta^{(0)}) -
e^{(0) \nu a} \gamma_{,\nu} e^{(0)}_{\ \ b \mu}
+ e^{(0) \nu}_{\ \ \ \ b} \gamma_{,\nu} e^{(0) a}_{\ \ \ \ \mu},
\\  \label{0.30}
\omega^{a}_{\ \ a_i m_j} = - \delta_{ij} e^{\phi^i - \gamma}
(e^{(0) a}_{\ \ \ \ \nu} \btd^{\nu}[g^{(0)}] \phi^i ) e^{(i)}_{\ \ a_i
m_i},
\\  \label{0.31}
\omega^{a_i}_{\ \ a m_j} = \delta_{ij} e^{\phi^i - \gamma}
(e^{(0) \nu }_{\ \ \ \ a} \p_{\nu} \phi^i ) e^{(i) a_i}_{\ \ \ \ m_i},
\\  \label{0.32}
\omega^{a_i}_{\ \ b_j m_k} = \delta_{ij} \delta_{jk}
\omega^{a_i}_{\ \ b_i m_i}(e^{(i)},\eta^{(i)}),
\ear
$i,j,k = 1, \ldots,n$, where $\omega^{a}_{\ \ b \mu}(e^{(0)},\eta^{(0)})$
and $\omega^{a_i}_{\ \ b_i m_i}(e^{(i)},\eta^{(i)})$,
are components  of the spin connections corresponding to the metrics from
(\ref{0.22}).

Let
\beq{0.33}
A_{M} \equiv \omega_{A B M} \hat{\Gamma}^A \hat{\Gamma}^B.
\eeq
For  $A_M = A_M(e,\eta,\hat{\Gamma})$ in the frame (\ref{0.21}) we get
\bear{0.34}
A_{\mu} = \omega^{(0)}_{a b \mu}  \hat{\Gamma}^a \hat{\Gamma}^b
+ (\Gamma_{\mu} \Gamma^{\nu} - \Gamma^{\nu} \Gamma_{\mu} ) \gamma_{, \nu},
\\  \label{0.35}
A_{m_i} = \omega^{(i)}_{a_i b_i m_i}  \hat{\Gamma}^{a_i}
\hat{\Gamma}^{b_i} + 2 \Gamma_{m_i} \Gamma^{\nu} \phi^{i}_{, \nu},
\ear
where $\omega^{(0)}_{a b \mu} = \omega_{a b \mu}(e^{(0)},\eta^{(0)})$
and $\omega^{(i)}_{a_i b_i m_i} = \omega_{a_i b_i m_i}(e^{(i)},\eta^{(i)})$,
$i =1, \ldots, n$.

\section{SUSY equations}

We consider the $D =11$ supergravity with the action
in the bosonic sector  \cite{CJS}
\beq{1.1}
S= \int d^{11}z \sqrt{|g|} \biggl\{R[g] - \frac{1}{4!} F^2 \biggr\}
+ c_{11} \int A \wedge F \wedge F,
\eeq
where $c_{11} = {\rm const}$ and  $F = d A$ is $4$-form.
Here we  consider pure bosonic configurations
in $D =11$ supergravity  (with zero fermionic fields)
that are solutions to the equations of motion corresponding
to the action (\ref{1.1}).

The number of supersymmetries (SUSY) corresponding to the
bosonic background $(e^{A}_M, A_{M_1 M_2 M_3})$
is defined by a dimension of the space of solutions
to (a set of) linear first-order differential equations
(SUSY eqs.)
\beq{1.2}
(D_M  + B_M ) \varepsilon = 0,
\eeq
where  $D_M$ is covariant spinorial derivative
>from (\ref{0.28}), $\varepsilon = \varepsilon (z)$  is
$32$-component "real" spinor field
(see Remark 1 below) and
\beq{1.3}
B_M  = \frac{1}{144 \sqrt{2}}
(\Gamma_M \Gamma^N \Gamma^P \Gamma^Q \Gamma^R -
12  \delta_M^N  \Gamma^P \Gamma^Q \Gamma^R) F_{NPQR}.
\eeq
Here  $F =  dA = \frac{1}{4!} F_{NPQR}
dz^{N} \wedge dz^{P} \wedge dz^{Q} \wedge dz^{R}$,
and $\Gamma_M$ are world $\Gamma$-matrices.

{\bf Remark 1.}
More  rigorously, $\varepsilon (z) \in \R^{0,32}_{{\bf G}} =
(\G_{\1})^{32}$, where $\G_{\1}$ is  an odd part of the 
infinite-dimensional Grassmann-Banach algebra (over $\R$) $\G = \G_0 
\oplus \G_1$ \cite{Iv}.

Here we consider the  decomposition  of matrix-valued  field
$B_M$  on the product manifold (\ref{0.10}) in the frame (\ref{0.21})
for electric and magnetic branes.

\subsection{$M2$-brane.}

Let the $4$-form be
\beq{2.1}
F = d \Phi \wedge \tau(I)
\eeq
where $\Phi = \Phi(x)$, $I = \{ i_1, \ldots, i_k \}$,
$i_1 < \ldots < i_k$, $d(I) =3$.
The calculations give
\bear{2.2}
B_{m_l} = \frac{1}{6 \sqrt{2}} s(I)
\exp(- \sum_{i \in I} d_i \phi^i )
[(1 - 3 \delta_I^l) \Gamma_{m_l} \Gamma^{\nu} \Phi_{, \nu}
- 3 \delta_0^l \Phi_{, m_l} ] \hat{\Gamma}(I),
\ear
where $l = 0, \ldots, n$, $m_0 = \mu$,
$s(I) = {\rm sign}(\prod_{i \in I} {\rm det}
(e^{(i) m_i}_{\ \ \ \ a_i}))$ and
$\hat{\Gamma}(I) =
\hat{\Gamma}^{\bar{1}} \hat{\Gamma}^{\bar{2}} \hat{\Gamma}^{\bar{3}}$
with $(\bar{1}, \bar{2}, \bar{3}) =
(1_{i_1}, \ldots, (d_{i_1})_{i_1}, \ldots,
1_{i_k}, \ldots, (d_{i_k})_{i_k})$.

\subsection{$M5$-brane.}
Let
\beq{2.3}
F =  (*_0 d \Phi) \wedge \tau(\bar{I}),
\eeq
where $*_0$ is the Hodge operator on $(M_0,g^0)$
and $\bar{I} = \{1, \dots, n \} \setminus I =
\{ j_1, \ldots, j_l \}$, $j_1 < \ldots < j_l$.
It follows from (\ref{2.3}) that $d_0 + d(\bar{I}) = 5$ and $d(I) = 6$.
We get
\bear{2.4}
B_{m_l} = \frac{1}{12 \sqrt{2}} s(\{ 0 \})
s(\bar{I})\exp[- (d_0 -2) \gamma -  \sum_{i \in \bar{I} } d_i
\phi^i ] \times \\ \nn
\times
[2 \Gamma_{m_l} \Gamma^{\nu} \Phi_{, \nu}
- 3 \delta_0^l (\Gamma_{m_l} \Gamma^{\nu}
- \Gamma^{\nu} \Gamma_{m_l}) \Phi_{, \nu}
+ 6 \delta_{\bar{I}}^l \Gamma^{\nu} \Gamma_{m_l}) \Phi_{, \nu}]
\hat{\Gamma}(\{ 0 \}) \hat{\Gamma}(\bar{I}),
\ear
where $l = 0, \ldots, n$; $s(\{0 \}) = {\rm sign}({\rm det}
(e^{(0) \nu}_{\ \ \ \ a}))$, $\hat{\Gamma}(\{0 \}) =
\hat{\Gamma}^{1_0} \ldots \hat{\Gamma}^{(d_0)_0}$
and $\hat{\Gamma}(\bar{I}) =
\hat{\Gamma}^{\bar{1}} \ldots \hat{\Gamma}^{\bar{k}}$
with $(\bar{1}, \ldots, \bar{k}) =
(1_{j_1}, \ldots, (d_{j_1})_{j_1}, \ldots,
1_{j_l}, \ldots, (d_{j_l})_{j_l})$.

\subsection{$2^{-k}$-splitting theorem}

In next section some examples of supersymmetric solutions
will be considered. In counting the fractional number of
supersymmetries the following ($2^{-k}$-splitting)
theorem is  used.

{\bf Theorem.}
{\em Let $V$ be  a vector space over $K = \R, \C$;
$V \neq \{0 \}$. Let $\Gamma_{[i]}: V \rightarrow V$,
$i =1, \ldots, k$, be a set of linear mappings (operators) satisfying:
\bear{2.5}
\Gamma_{[i]}^2 = {\rm id}_V \equiv \1, \qquad
\Gamma_{[i]} \circ \Gamma_{[j]} =
\Gamma_{[j]} \circ \Gamma_{[i]},
\ear
$i,j = 1, \ldots, k$. Then
\bear{2.6}
V = \oplus \sum_{s_1, \dots, s_k = \pm 1} V_{s_1, \dots, s_k},
\ear
where
\bear{2.7}
V_{s_1, \dots, s_k} \equiv \{x \in V| \Gamma_{[i]} x = s_i x,
i =1, \ldots, k \},
\ear
are subspaces of $V$, $s_1, \dots, s_k = \pm 1$. Moreover, if there
exists a set of linear bijective mappings
$A_{[i]}: V \rightarrow V$, $i =1, \ldots, k$,  satisfying
\bear{2.8}
A_{[i]} \circ \Gamma_{[i]} = - \Gamma_{[i]} \circ A_{[i]},
\\ \nn \label{2.9}
A_{[i]} \circ \Gamma_{[j]} =  \Gamma_{[j]} \circ A_{[i]},
\quad i \neq j,
\ear
$i,j = 1, \ldots, k$, then all subspaces $V_{s_1, \dots, s_k}$
are mutually isomorphic and for finite-dimensional $V$
\bear{2.10}
{\rm dim} V_{s_1, \dots, s_k} = 2^{-k} {\rm dim} V,
\ear
$s_1, \dots, s_k = \pm 1$. }

{\bf Proof.}
Let us introduce a set of projector operators
\bear{2.11}
P_{s}^{[i]} = \frac{1}{2} (\1 + s \Gamma_{[i]}),
\ear
$i =1, \ldots, k$; $s = \pm 1$, satisfying
\bear{2.12}
(P_{s}^{[i]})^2 = P_{s}^{[i]},
\\  \label{2.13}
P_{s}^{[i]} + P_{-s}^{[i]} = \1,
\\  \label{2.14}
P_{s}^{[i]} \circ P_{-s}^{[i]} = \0,
\\  \label{2.15}
P_{s}^{[i]} \circ P_{s'}^{[j]} = P_{s}^{[j]} \circ P_{s'}^{[i]},
\ear
for all $i,j =1, \ldots, k$; $s, s' = \pm 1$.
$\0$ is the zero-operator. The relation (\ref{2.13}) implies
\bear{2.16}
(P_{+1}^{[1]} + P_{-1}^{[1]}) \circ \ldots \circ
(P_{+1}^{[k]} + P_{-1}^{[k]}) =
\sum_{s_1, \dots, s_k = \pm 1}
P_{s_1}^{[1]} \circ \ldots  \circ P_{s_k}^{[k]}.
\ear
By definition
\bear{2.17}
V_{s_1, \dots, s_k} =
{\rm Ker} P_{-s_1}^{[1]} \cap \ldots \cap {\rm Ker} P_{-s_k}^{[k]}.
\ear
It may be verified using (\ref{2.11})-(\ref{2.16})
that
\bear{2.18}
V_{s_1, \dots, s_k} =
 (P_{s_1}^{[1]} \circ  \ldots  \circ P_{s_k}^{[k]}) V
\ear
and the decomposition (\ref{2.6}) holds.
>From (\ref{2.8}) and (\ref{2.11}) we get
\bear{2.20}
A_{[i]} \circ P_{s}^{[i]} =  P_{-s}^{[i]} \circ A_{[i]},
\\ \label{2.21}
A_{[i]} \circ P_{s}^{[j]} =  P_{s}^{[j]} \circ A_{[i]}.
\qquad i \neq j,
\ear
$i,j =1, \ldots, k$; $s = \pm 1$.
Let us introduce linear functions
\bear{2.22}
A_{s_1, \dots, s_k}:
V_{s_1, \dots, s_i, \dots, s_k} \rightarrow
V_{s_1, \dots, -s_i, \dots, s_k}
\\ \nn
v \mapsto A_{[i]} v.
\ear
It follows from
(\ref{2.18}), (\ref{2.20}) and (\ref{2.21}) that these
functions are correctly defined and are bijective ones.
This implies that all subspaces $V_{s_1, \dots, s_k}$
are mutually isomorphic. The Theorem is proved.

We note that  projector operators (\ref{2.11})
with $\Gamma_{[i]}$ being a product of  $\Gamma$-matrices
were considered previously in \cite{AIR}.

\section{Examples of supersymmetric solutions}

\subsection{$M2$-brane.}

We consider the electric  $2$-brane solution
defined on  the manifold
\beq{3.2a}
M_{0}  \times M_{1}  \times M_{2}.
\eeq
The solution reads
\bear{3.2}
g= H^{1/3} \{ \hat{g}^0  +  H^{-1} \hat{g}^1 +  \hat{g}^2 \},
\\ \label{3.3}
F = \nu d H^{-1} \wedge \hat{\tau}_1,
\ear
where  $\nu^2 = 1/2$, $H = H(x)$ is a harmonic function on $(M_0,g^0)$
$d_1 = 3$, $d_0 + d_2 = 8$, and the metrics $g^i$, $i = 0,1,2$, 
are Ricci-flat.

\subsubsection{Flat $g^i$}

Let us consider a special case of flat $g^i$
\beq{3.6}
g^{0}_{\mu \nu} = \delta_{\mu \nu}, \quad
g^{1}_{m_1 n_1} = \eta^{(1)}_{m_1 n_1}, \quad
g^{2}_{m_2 n_2} = \delta_{m_2 n_2}
\eeq
where $(\eta^{(1)}_{a_1 b_1}) = {\rm diag}(-1,+1,+1)$.
We fix the frames in (\ref{0.21}) as follows
\beq{3.7}
e^{(0) a}_{\ \ \ \ \mu} = \delta^{a}_{\nu}, \qquad
e^{(i) a_i}_{\ \ \ \ m_i} = \delta^{a_i}_{m_i},
\eeq
$i = 1,2$.

It may be verified using relations from subsections 2.3 and 2.4
and formulae (\ref{3.6}) and  (\ref{3.7}) that the SUSY eqs.
(\ref{1.2}) are satisfied identically if
\bear{3.8}
\varepsilon  = H^{-1/6} \varepsilon_{*}, \qquad
\varepsilon_{*} = {\const},
\\ \label{3.9}
\Gamma \varepsilon_{*} = c \varepsilon_{*},
\qquad c = {\rm sign} \nu,
\ear
where
\beq{3.10}
\Gamma = \hat{\Gamma}^{1_1} \hat{\Gamma}^{2_1} \hat{\Gamma}^{3_1}.
\eeq
Here  $\Gamma$ is real-valued matrix satisfying
$\Gamma^2 = {\bf 1}$,  where ${\bf 1}$ is unit
$32 \times 32$-matrix. Let $A = \hat{\Gamma}^{1_0}$.
The pair $\Gamma = \Gamma_{[1]}, A = A_{[1]}$ satisfies the conditions of
the {\bf Theorem}, and hence for  $\varepsilon_1 \in \R^{32}$
the dimension of the subspace
$V_c$  of solutions to eqs. (\ref{3.9}) is $16$.
For   $\varepsilon_1 \in \G_1^{32}$
(see {\bf Remark 1}) the (odd part of)
superdimension of the subsuperspace ${\bf V_c}$ from (\ref{3.9}) is
also $16$.  This means that (at least) $N =1/2$ part of SUSY is preserved.

\subsubsection{Non-flat $g^i$.}

Here we put $d_2 =0$ in (\ref{3.2a}), i.e. we
consider the metric  on  $M_{0}  \times M_{1}$:
\beq{3.2n}
g= H^{1/3} \{ \hat{g}^0  +  H^{-1} \hat{g}^1 \},
\eeq
with $d_1 = 3$, $d_0 = 8$, and the form from (\ref{3.3})
where metrics $g^i$, $i = 0,1$ are Ricci-flat;
$g^0$ has Euclidean signature and $g^1$ has
the signature ${\rm diag}(-1,+1,+1)$.

Let us consider $\Gamma$-matrices
\beq{3.3n}
(\hat{\Gamma}^A) = (\hat{\Gamma}^{a}_{(0)} \otimes \1_2,
\hat{\Gamma}_{(0)} \otimes \hat{\Gamma}^{a_1}_{(1)}),
\eeq
where $\hat{\Gamma}^{a}_{(0)}$, $a = 1_0, \ldots, 8_0$
correspond to $M_0$ and
$\hat{\Gamma}^{a_1}_{(1)}$, $a_1 = 1_1, 2_1, 3_1$ correspond
to $M_1$ and $\hat{\Gamma}_{(0)} = \hat{\Gamma}^{1_0}_{(0)}
\ldots \hat{\Gamma}^{8_0}_{(0)}$.
The substitution
\bear{3.8n}
\varepsilon  = H^{-1/6} \eta_0(x) \otimes \eta_1(y), \qquad
\\ \label{3.9n}
\Gamma \varepsilon = c_{\nu} \varepsilon,
\qquad c = {\rm sign} \nu,
\ear
where $\Gamma$ is defined in  (\ref{3.10}),
$\eta_0(x)$ is a $2$-component
Killing spinor on $M_0$ and $\eta_1(y)$ is
a $16$-component Killing spinor on $M_1$, i.e.
\beq{3.8k}
D^{(0)}_{\mu} \eta_0 = D^{(1)}_{m_1} \eta_1 = 0,
\eeq
with $D^{(0)}_{\mu} = \p_{\mu} +  \frac{1}{4} \omega^{(0)}_{a b \mu}
\hat{\Gamma}^a_{(0)} \hat{\Gamma}^b_{(0)}$
and $D^{(1)}_{m_1} = \p_{m_1} +  \frac{1}{4} \omega^{(1)}_{a_1 b_1
m_1} \hat{\Gamma}^{a_1}_{(1)} \hat{\Gamma}^{b_1}_{(1)}$,
gives us a solution to the SUSY equations.

We get from  (\ref{3.3n})  that
\beq{3.9k}
\Gamma = \hat{\Gamma}_{(0)} \otimes \hat{\Gamma}_{(1)},
\eeq
where $\hat{\Gamma}_{(1)} = \hat{\Gamma}^{1_1}_{(1)}
\hat{\Gamma}^{2_1}_{(1)} \hat{\Gamma}^{3_1}_{(1)}$.
Choosing real matrices
$\hat{\Gamma}^{1_1}_{(1)} = i \sigma_2$,
$\hat{\Gamma}^{2_1}_{(1)} =  \sigma_1$,
$\hat{\Gamma}^{3_1}_{(1)} =  \sigma_3$,
(where $\sigma_i$ are the standard Pauli matrices)
we get $\hat{\Gamma}_{(1)} =  \1_2$, and due to
eq. (\ref{3.9k})  the relation (\ref{3.9n})
is equivalent to the following
\beq{3.9nk}
\hat{\Gamma}_{(0)} \eta_{(0)} = c \eta_{(0)}.
\eeq
Hence the number of unbroken SUSY is (at least)
\beq{3.10nk}
N = n_0(c) n_1/32,
\eeq
where  $n_0(c)$ is the number of
chiral Killing spinors on $M_0$ satisfying
$(\ref{3.9nk})$ with $c = {\rm sign} \nu$,  and
$n_1$  is the number of Killing spinors on $M_1$.

\subsection{$M5$-brane}

Now let us consider the magnetic  $5$-brane solution
defined on the manifold  (\ref{3.2a}),
\bear{3.4m}
g= H^{2/3} \{ \hat{g}^0  +  H^{-1} \hat{g}^1 +  \hat{g}^2 \},
\\ \label{3.5m}
F = \nu (*_0 d H) \wedge \hat{\tau}_2,
\ear
where $\nu^2 = 1/2$,  $H = H(x)$ is a harmonic function on $(M_0,g^0)$,
$d_1 = 6$, $d_0 + d_2 =5$ and metrics $g^i$, $i = 0,1,2$, are Ricci-flat.

\subsubsection{Flat $g^i$}

Let all metrics be flat, i.e. we consider
the relations (\ref{3.6}) with $(\eta^{(1)}_{a_1 b_1}) =
\\ {\rm diag}(-1,+1,+1,+1,+1,+1)$.
We also consider canonical frames defined by (\ref{3.7}).

The SUSY eqs. (\ref{1.2}) are satisfied identically if
\beq{3.8m}
\varepsilon  = H^{-1/12} \varepsilon_{*},
\qquad \varepsilon_{*} = {\const},
\eeq
where $\varepsilon_{*}$ obeys to  eq. (\ref{3.9}) with
\beq{3.10m}
\Gamma = \hat{\Gamma}^{\bar{1}} \hat{\Gamma}^{\bar{2}}
\hat{\Gamma}^{\bar{3}} \hat{\Gamma}^{\bar{4}} \hat{\Gamma}^{\bar{5}}
\eeq
and
$(\bar{1}, \ldots, \bar{5}) =
(1_{0}, \ldots, (d_{0})_{0}, 1_{2}, \ldots, (d_{2})_{2})$.
Here  $\Gamma^2 = {\bf 1}$. Let $A = \hat{\Gamma}^{1_1}$.
The pair $\Gamma = \Gamma_{[1]}, A = A_{[1]}$ satisfies the conditions of
the {\bf Theorem}. Hence  we obtain  that $N =1/2$ part of supersymmetries
is preserved.

\subsubsection{Non-flat $g^i$.}

Let $d_2 =0$ in (\ref{3.4m}), i.e. we
consider the metric  on  $M_{0}  \times M_{1}$
\beq{3.4mn}
g= H^{2/3} \{ \hat{g}^0  +  H^{-1} \hat{g}^1 \},
\eeq
with $d_1 = 6$, $d_0 = 5$, where metrics $g^i$, $i = 0,1$, are Ricci-flat,
$g^0$ has a Euclidean signature and $g^1$ has
the signature ${\rm diag}(-1,+1,+1,+1,+1,+1)$.
The 4-form (\ref{3.5m}) is modified as follows
\bear{3.5mn}
F = \nu (*_0 d H).
\ear

Let us consider $\Gamma$-matrices
\beq{3.3nm}
(\hat{\Gamma}^A) = (\hat{\Gamma}^{a}_{(0)} \otimes
\hat{\Gamma}_{(1)}, \1_4 \otimes \hat{\Gamma}^{a_1}_{(1)}),
\eeq
where $\hat{\Gamma}^{a}_{(0)}$, $a = 1_0, \ldots, 5_0$,
correspond to $M_0$ and
$\hat{\Gamma}^{a_1}_{(1)}$, $a_1 = 1_1, \ldots, 6_1$,  correspond
to $M_1$ and $\hat{\Gamma}_{(1)} = \hat{\Gamma}^{1_1}_{(1)}
\ldots \hat{\Gamma}^{6_1}_{(1)}$.
The substitution
\bear{3.8nm}
\varepsilon  = H^{-1/12} \eta_0(x) \otimes \eta_1(y), \qquad
\\ \label{3.9nm}
\Gamma \varepsilon = c_{\nu} \varepsilon,
\qquad c = {\rm sign} \nu,
\ear
where $\Gamma$ from  (\ref{3.10m}) reads
\beq{3.10mn}
\Gamma = \hat{\Gamma}^{1_{0}} \ldots \hat{\Gamma}^{5_{0}}
= \hat{\Gamma}_{(0)} \otimes \hat{\Gamma}_{(1)},
\eeq
with $\hat{\Gamma}_{(0)} = \hat{\Gamma}^{1_0}_{(0)}
\ldots \hat{\Gamma}^{5_0}_{(0)}$
gives us a solution to the SUSY equations.
In (\ref{3.8nm}) $\eta_0(x)$ is $4$-component
Killing spinor on $M_0$ and $\eta_1(y)$ is
$8$-component Killing spinor on $M_1$, i.e.
the relations (\ref{3.8k}) are  satisfied.
We put
$\hat{\Gamma}^{5_0}_{(0)} =
\hat{\Gamma}^{1_0}_{(0)} \ldots \hat{\Gamma}^{4_0}_{(0)}$.
Then $\hat{\Gamma}_{(0)} =  \1_4$, and due to
(\ref{3.10mn})  the relation (\ref{3.9nm}) reads
\beq{3.9nkm}
\hat{\Gamma}_{(1)} \eta_{(1)} = c \eta_{(1)}.
\eeq
Hence, the number of  preserved SUSY is (at least)
\beq{3.10nkm}
N = n_0 n_1(c)/32,
\eeq
where  $n_1(c)$ is the number of
(chiral) Killing spinors on $M_1$ satisfying
(\ref{3.9nkm}) with $c = {\rm sign} \nu$,  and
$n_0$  is the number of Killing spinors on $M_0$.
A special case of this supersymmetric solution
with $M_0 = \R^5$  and $M_1 = \R^2 \times K3$,
was considered in \cite{K2}. In this case
$N =1/4$ in agreement with  (\ref{3.10nkm})
since $n_0 = 4$ and (as can be easily verified) 
$n_1(c) = n[K3] =2$ (i.e. the 
number of chiral Killing spinors on $\R^2 \times K3$ is equal to the total 
number of Killing spinors on $K3$).  We remind that $K3$  
is a 4-dimensional  Ricci-flat K\"{a}hler manifold with 
$SU(2)$ holonomy group and self-dual (or anti-self-dual) curvature 
tensor.  $K3$ has two Killing spinors (left or right).

\subsection{$M2 \cap M5$-branes}

Here we consider solutions  with two "orthogonally" intersecting
$p$-branes  (with $p =2,5$) defined on the manifold
\bear{4.1}
M_{0}  \times M_{1}  \times M_{2} \times M_{3} \times M_{4}
\ear
to show how the Theorem  works.

The solution  with  $M2$ and $M5$ branes
defined on the manifold (\ref{4.1}) reads  \cite{PT,TsO,GKT,IM0}
\bear{4.2}
g= H_1^{1/3} H_2^{2/3} \{ \hat{g}^0  +
H_1^{-1} \hat{g}^1 +  H_2^{-1} \hat{g}^2 +
H_1^{-1} H_2^{-1} \hat{g}^3 +  \hat{g}^4 \},
\\ \label{4.3}
F = \nu_1 d H^{-1}_1 \wedge \hat{\tau}_1 \wedge \hat{\tau}_3 +
\nu_2 (*_0 d H_2) \wedge  \hat{\tau}_1 \wedge \hat{\tau}_4,
\ear
where $\nu^2_1 = \nu^2_2 = 1/2$; $H_1, H_2$ are harmonic
functions on $(M_0,g^0)$, $d_1 =1$, $d_2 =4$, $d_3 = 2$, $d_0 +d_4 =4$,
and metrics $g^i$, $i = 0,1,2,3,4$, are Ricci-flat.

Let  all $g^i$ be flat:
\beq{4.4}
g^{0}_{\mu \nu} = \delta_{\mu \nu}, \quad
g^{3}_{m_1 n_1} = \eta^{(3)}_{m_1 n_1}, \quad
g^{i}_{m_i n_i} = \delta_{m_i n_i}, \quad i = 1, 2, 4,
\eeq
where $(\eta^{(3)}_{a_1 b_1}) = {\rm diag}(-1,+1)$.
We consider the frames from (\ref{3.7}) with
$i =1,2,3,4$.

The SUSY eqs. (\ref{1.2}) are satisfied identically if
\bear{4.5}
\varepsilon  = H_1^{-1/6} H_2^{-1/12} \varepsilon_{*}, \qquad
\varepsilon_{*} = {\const},
\\ \label{4.6}
\Gamma_{[i]} \varepsilon_{*} = c_i \varepsilon_{*},
\qquad c_i  = {\rm sign} \nu_i,
\ear
$i =1,2$, where
\bear{4.7}
\Gamma_{[1]} = \hat{\Gamma}^{1_1} \hat{\Gamma}^{1_3} \hat{\Gamma}^{2_3},
\qquad
\Gamma_{[2]} = \hat{\Gamma}^{\bar{1}} \hat{\Gamma}^{\bar{2}}
\hat{\Gamma}^{\bar{3}} \hat{\Gamma}^{\bar{4}} \hat{\Gamma}^{\bar{5}},
\ear
$(\bar{1}, \ldots, \bar{5}) =
(1_{0}, \ldots, (d_{0})_{0}, 1_1, 1_{4}, \ldots, (d_{4})_{4})$.

Introducing the "complimentary" matrices
$A_{[1]} = \hat{\Gamma}^{1_0}$ and $A_{[2]} = \hat{\Gamma}^{1_3}$,
we get from the  {\bf Theorem} that  the (super)dimension of the
(super)subspace of solutions to eqs. (\ref{4.6}) is $8$, i.e.  
at least $N =1/4$ part of SUSY "survives".

{\bf Remark 2.}  The configurations under
consideration  remain supersymmetric if
the functions $H_i$ are arbitrary (not obviously harmonic ones).
Thus, we are led to supersymmetric field sets
that are not solutions to the equations of motion.

\section{Conclusions}

Thus here we considered
the "Killing-like"  SUSY equations for $D=11$ supergravity
in the backgrounds with a block-diagonal metric
defined on  the product of Ricci-flat spaces
$M_{0}  \times M_{1} \times \ldots \times M_{n}$.

We obtained decomposition relations for
ingredients of the
"Killing-like"  SUSY equations (e.g. spin connection, matrix-valued
covector field) and proved the $2^{-k}$-splitting theorem
for $k$ commuting linear operators.
We considered  examples of $M2$ and $M5$ branes
defined on the product  of two Ricci-flat spaces and obtained
formulae  for a fractional number of unbroken SUSY.
Also $p$-brane
$M2 \cap M5$-configuration  defined on the product of four flat spaces
is considered to illustrate how the "$2^{-k}$-splitting" theorem
works.

Other examples (with several branes on the products of Ricci-flat spaces
) in different supergravity models will be considered
in a separate publications.

\begin{center}
{\bf Acknowledgments}
\end{center}

This work was supported in part by DFG grant 436 RUS 113/236/O(R)
and by the Russian Ministry for  Science and Technology and Russian Fund 
for Basic Research, grant N 98-02-16414.


  \end{document}